\documentstyle[aps,prd,preprint,epsfig]{revtex}

\newcommand{\beq}{\begin{equation}}
\newcommand{\eeq}{\end{equation}}
\newcommand{\beqr}{\begin{eqnarray}}
\newcommand{\eeqr}{\end{eqnarray}}

\begin{document}

\preprint{LPT-ORSAY 03-64}
\draft
\title{Inflating magnetically charged braneworlds}

\author{Inyong Cho\footnote{Electronic address: 
Inyong.Cho@th.u-psud.fr}}
\address{Laboratoire de Physique Th\'eorique,
Universit\'e Paris-Sud, B\^atiment 210, F-91405   
Orsay CEDEX, France}
\author{Alexander Vilenkin\footnote{Electronic address: 
vilenkin@cosmos.phy.tufts.edu}}
\address{Institute of Cosmology, Department of Physics and Astronomy,
Tufts University, Medford, Massachusetts 02155, USA}

\date{\today}

\maketitle

\begin{abstract}
Numerical solutions of Einstein, scalar, and gauge field equations are
found for static and inflating defects in a higher-dimensional
spacetime. The defects have $(3+1)$-dimensional core and magnetic
monopole configuration in $n=3$ extra dimensions. For symmetry-breaking 
scale $\eta$ below the critical value $\eta_c$, the defects
are characterized by a flat worldsheet geometry and asymptotically
flat extra dimensions. The critical scale $\eta_c$ is comparable to
the higher-dimensional Planck scale and has some dependence on the
gauge and scalar couplings.  For $\eta=\eta_c$, the extra dimensions
degenerate into a `cigar', and for $\eta>\eta_c$ all static solutions
are singular. The singularity can be removed if the requirement of
staticity is relaxed and defect cores are allowed to inflate. The
inflating solutions have de Sitter worldsheets and cigar geometry in
the extra dimensions. Exact analytic solutions describing the
asymptotic behavior of these inflating monopoles are found and the
parameter space of these solutions is analyzed. 
\end{abstract}

\vspace{0.5in}

\section{Introduction}

In ``braneworld'' models, our universe is represented by a
$(3+1)$-dimensional brane floating in a higher-dimensional bulk
spacetime~\cite{Rubakov,Arkani,RS}. The brane can be thought of as
fundamental $D$-brane, or it can arise as a topological defect in a
higher-dimensional field theory.  It may well be that $D$-branes could
also eventually find some dual field-theory description.  In the
simplest, codimension-one models, the brane can be pictured as a
domain wall propagating in a 5D spacetime~\cite{Rubakov,Wall}.
Higher codimensions with both gauge and global defects have also been
considered.  In particular, models have been discussed where the field
configuration in the directions orthogonal to the brane is that of a
cosmic string~\cite{String}, a monopole~\cite{Olasagasti,Monopole}, or
a texture~\cite{Texture}.  In these models, the bulk curvature produced
by the defects plays a major role in localizing gravity on the brane
and in solving the mass-hierarchy problem.

The physics of braneworld models crucially depends on the nature of
the gravitational fields produced by defects in higher
dimensions. These fields, however, are not yet fully understood.
Exact analytic solutions have been obtained to the higher-dimensional
Einstein equations, and to the combined system of Einstein, gauge
and/or Goldstone field equations for branes carrying gauge or global
charges \cite{RS,Gregorypbrane,Olasagasti,Monopole,GM}. These
solutions correspond to the limit of a vanishingly thin core and
contain curvature singularities. One can hope that the singularities
will be smoothed out in more realistic models including the defect
core, and it has been verified in
\cite{GM} that this is indeed the case for higher-dimensional global
monopoles.

Apart from the spurious singularities resulting from the thin-core
approximation, the defect spacetimes can also develop singularities in
the exterior region. Well-known $3$-dimensional examples are static
domain wall \cite{AV81} and global string \cite{stringsing}
solutions. In these cases, the singularities are removed by relaxing
the requirement of staticity and allowing the defect worldsheets to
inflate \cite{AV83,Gregory}.  Static 3-dimensional gauge strings are
also known to be singular when their symmetry-breaking scale becomes
greater than the Planck scale (see, e.g., Ref.~\cite{V94} for
discussion and references).

In Ref.~\cite{GM} we showed that similar singularities arise in
higher-dimensional monopole spacetimes, if the symmetry-breaking scale
of the monopole exceeds certain critical value comparable to the
Planck scale, and that the singularities can be removed if the
monopole worldsheets are allowed to inflate. We argued that the
inflation rate $H$ is uniquely fixed by the regularity condition.

Another motivation for looking into the super-Planckian regime is the
interesting work by Dvali, Gabadadze and Shifman \cite{DGS},
suggesting that braneworld models of codimension $n>2$, with a
very low bulk Planck scale, can provide a solution to the cosmological
constant problem. It was conjectured in Ref.~\cite{DGS} that defect
solutions in such models may exhibit de Sitter inflation of the
worldsheet, with the expansion rate {\it inversely} proportional to
the brane tension. The very low expansion rate observed today could
then be obtained with a very large brane tension. Our analysis of
global defects in \cite{GM} has shown that global higher-dimensional
defects do not have this property.

The purpose of the present paper is to extend the analysis of
Ref.~\cite{GM} to the case of gauge defects. Specifically, we consider
a codimension-3 gauge monopole. In the following Section, we introduce
the model and review some asymptotic solutions found earlier by
Gregory \cite{Gregorypbrane,Gregorypinf}. The numerical solutions
assuming a static (noninflating) worldsheet are presented in Section
III, for both sub- and super-critical regimes. We find that
sub-critical solutions are regular and asymptotically flat, while the
super-critical ones have a curvature singularity at a finite distance
from the core (as they do in the global monopole case).  Inflating
defect spacetimes are discussed in Sec.~IV. We find analytic solutions
describing the asymptotic behavior of these spacetimes and chart the
rather complicated parameter space of these solutions. Our numerical
results indicate that inflation of the worldsheet does remove the
singularity and that the resulting spacetimes have a `cigar' geometry.
However, the numerical accuracy limitations do not allow us to
conclude that the regularity requirement selects a unique value of the
expansion rate $H$.  Our conclusions are briefly summarized in Section
V.

\section{Field equations and some asymptotic solutions}

The action for our model is
\begin{equation}
{\cal S} = \int d^7x \sqrt{-g}\left(
{{\cal R} \over  2\kappa^2} 
+{\cal L}_m
\right)\,,
\label{eq=action}
\end{equation}
where $\kappa^2=1/M^{2+n}$ and $M$ is the 7D Planck mass.
We assume that the metric has de Sitter or Poincare symmetry along the
brane and spherical symmetry in the extra three dimensions.  This
corresponds to the following {\it ansatz}:
\beq
d s^2 = B^2(r)d\bar{s}_4^2+ dr^2 + C^2(r)r^2
(d\theta^2+\sin^2\theta d\varphi^2)\,,
\label{eq=metric}
\eeq
where $d\bar{s}_4^2$ is the 4D world-volume metric.  

The matter-field Lagrangian is given by
\beq
{\cal L}_m = -{1\over 4}G^a_{MN}G^{aMN}
-{1\over 2}D_M\Phi^aD^M\Phi^a
-{\lambda\over 4}(\Phi^a\Phi^a -\eta^2)^2\,,
\eeq
where
\begin{eqnarray}
D_M\Phi^a &=& \partial_M\Phi^a + e\epsilon^{abc}W^b_M\Phi^c\,,\\
G^a_{MN} &=& \partial_MW^a_N-\partial_NW^a_M
+ e\epsilon^{abc}W^b_MW^c_N\,,
\end{eqnarray}
$D_M$ is the covariant derivative,
and $e$ is the gauge-coupling constant.\footnote{Here, small Roman indices 
from the beginning of the alphabet run through the values
$a,b,c=1,2,3$, and from the middle of the alphabet $i,j=4,5,6$. 
Capital Roman indices take values from 0 to 6, and 
Greek indices from 0 to 3.}
In the following sections, it will be convenient to use the notation
\beq
\beta\equiv\lambda/e^2
\eeq
for the ratio of the scalar and gauge couplings.

We assume that the scalar triplet $\Phi^a$  has a ``hedgehog''
configuration in extra dimensions,
\beq
\Phi^a = \phi (r){x^{3+a}\over{r}}\,.
\label{eq=Phi}
\eeq
The gauge field is given by
\begin{eqnarray}
W^a_\mu &=& 0\,,\\ 
W^a_i &=& [1-w(r)]\epsilon^{a(i-3)(j-3)}{x^j\over er^2}\,,\label{eq=W}
\end{eqnarray}
or, in spherical coordinates (\ref{eq=metric}),
\begin{eqnarray}
W^a_r &=& 0\,,\\
W^a_\theta &=& {1-w(r)\over er}(\sin\varphi,-\cos\varphi,0)\,,\\
W^a_\varphi &=& {1-w(r)\over er}\sin\theta 
(\cos\theta\cos\varphi,\cos\theta\sin\varphi,-\sin\theta)\,.
\end{eqnarray}

Einstein equations corresponding to the above {\it ans\"atze} are
\begin{eqnarray}
-G^\mu_\mu &=& -3{B'' \over B} -3\left({B'\over B}\right)^2
-6{B' \over Br} -6{B'C'\over BC} -2{C''\over C}
-\left({C'\over C}\right)^2 -6{C'\over Cr}
+{1\over C^2r^2}-{1\over r^2}+{1\over 4}{\bar{R}^{(4)}\over B^2} 
\nonumber \\
{} &=& \kappa^2\left[ {\phi'^2 \over 2} +{\phi^2w^2\over C^2r^2}
+{1\over e^2C^2r^2}\left[ w'^2
+ {(1-w^2)^2\over 2C^2r^2} \right]
+{\lambda\over 4}(\phi^2-\eta^2)^2
\right] \,,
\label{eq=Gtt} \\
-G^r_r &=& -6\left({B'\over B}\right)^2
-8{B' \over Br} -8{B'C'\over BC}
-\left({C'\over C}\right)^2 -2{C'\over Cr}
+{1\over C^2r^2}-{1\over r^2}+{1\over 2}{\bar{R}^{(4)}\over B^2} 
\nonumber \\
{} &=& \kappa^2\left[ -{\phi'^2 \over 2} +{\phi^2w^2 \over C^2r^2}
+{1\over e^2C^2r^2}\left[ -w'^2
+ {(1-w^2)^2\over 2C^2r^2} \right] 
+{\lambda\over 4}(\phi^2-\eta^2)^2
\right] \,,
\label{eq=Grr}\\
-G^{\theta_i}_{\theta_i} &=& 
-4{B'' \over B} -6\left({B'\over B}\right)^2
-4{B' \over Br} -4{B'C'\over BC} -{C''\over C}
-2{C'\over Cr}
+{1\over 2}{\bar{R}^{(4)}\over B^2} 
\nonumber \\
{} &=& \kappa^2\left[ {\phi'^2 \over 2}
-{(1-w^2)^2 \over 2e^2C^4r^4}
+{\lambda\over 4}(\phi^2-\eta^2)^2
\right]\,, 
\label{eq=Gthth}
\end{eqnarray}
where $\bar{R}^{(4)}$ is the Ricci scalar of the 4D world volume.

The field equation for the scalar field is
\begin{equation}
\phi'' + 2\left( 2{B'\over B} +{C'\over C}+{1\over r}\right)\phi'
-2{w^2\phi \over C^2r^2} -\lambda\phi (\phi^2-\eta^2) =0\,.
\label{eq=phi}
\end{equation}
The field equation for the gauge field is
\beq
w''+4{B'\over B}w' +{w(1-w^2)\over C^2r^2}-e^2\phi^2w=0\,.
\label{eq=w}
\eeq

At large distances from the monopole core, $r\gg
(\sqrt{\lambda}\eta)^{-1}$, one can expect the scalar and gauge field
to approach the asymptotic values
\beq
\phi(r\to\infty)=\eta,\quad w(r\to\infty)=0.
\label{asympt}
\eeq
In this regime, the monopole can be approximated as an infinitely thin
charged 3-brane. The corresponding solution of Einstein equations
has been found by Gregory \cite{Gregorypbrane} in the case of a flat
4D world-volume metric. We shall not reproduce this solution here
and note only that it is asymptotically flat,
$B(r\to\infty)=C(r\to\infty)=1$.  

Gregory has also considered inflating branes with a de Sitter
world-volume metric~\cite{Gregorypinf},
\begin{equation}
d\bar{s}_4^2 = -dt^2 +H^{-2}\cosh^2(Ht)d\Omega_3^2\,.
\label{deSitter}
\end{equation}
Using higher-dimensional vacuum Einstein's equations, she found
the asymptotic behavior of the metric at large $r$. In the case of a
3-brane in a 7D spacetime, her asymptotic solution has the form,
\beq
ds^2 = {3\over 5}H^2r^2 (-dt^2 +e^{2Ht}d\text{x}^2)
+dr^2 +{1\over 5}r^2d\Omega_2^2\,.
\label{Gregory}
\eeq 
Note that this is not globally flat at $r\to\infty$, but has a
solid deficit angle in the extra dimensions. Although this asymptotic
solution was obtained for an uncharged brane, one can expect it to
give a good approximation, since the gauge field energy-momentum
tensor vanishes at $r\to\infty$.
 
In the following sections, we shall see that Gregory's solutions
indeed describe the asymptotic behavior of the metric, but only if the
symmetry-breaking scale $\eta$ is sufficiently small.  For large
values of $\eta$, the nature of the solutions is drastically
different.

\section{Static solutions}

We numerically solved Einstein, scalar-field, and gauge-field equations
(\ref{eq=Gtt})-(\ref{eq=w}) with boundary conditions, $\phi (0)=0$,
$\phi(\infty)=\eta$, $w(0)=1$, $w(\infty)=0$, $B(0)=C(0)=1$, and
$B'(0)=C'(0)=0$~\cite{numerics}.  
For static solutions, we assumed a flat worldsheet,
$d\bar{s}_4^2 = \eta_{\mu\nu}dx^\mu dx^\nu$, and $\bar{R}^{(4)}=0$.
 
\begin{figure}
\begin{center}
\epsfig{file=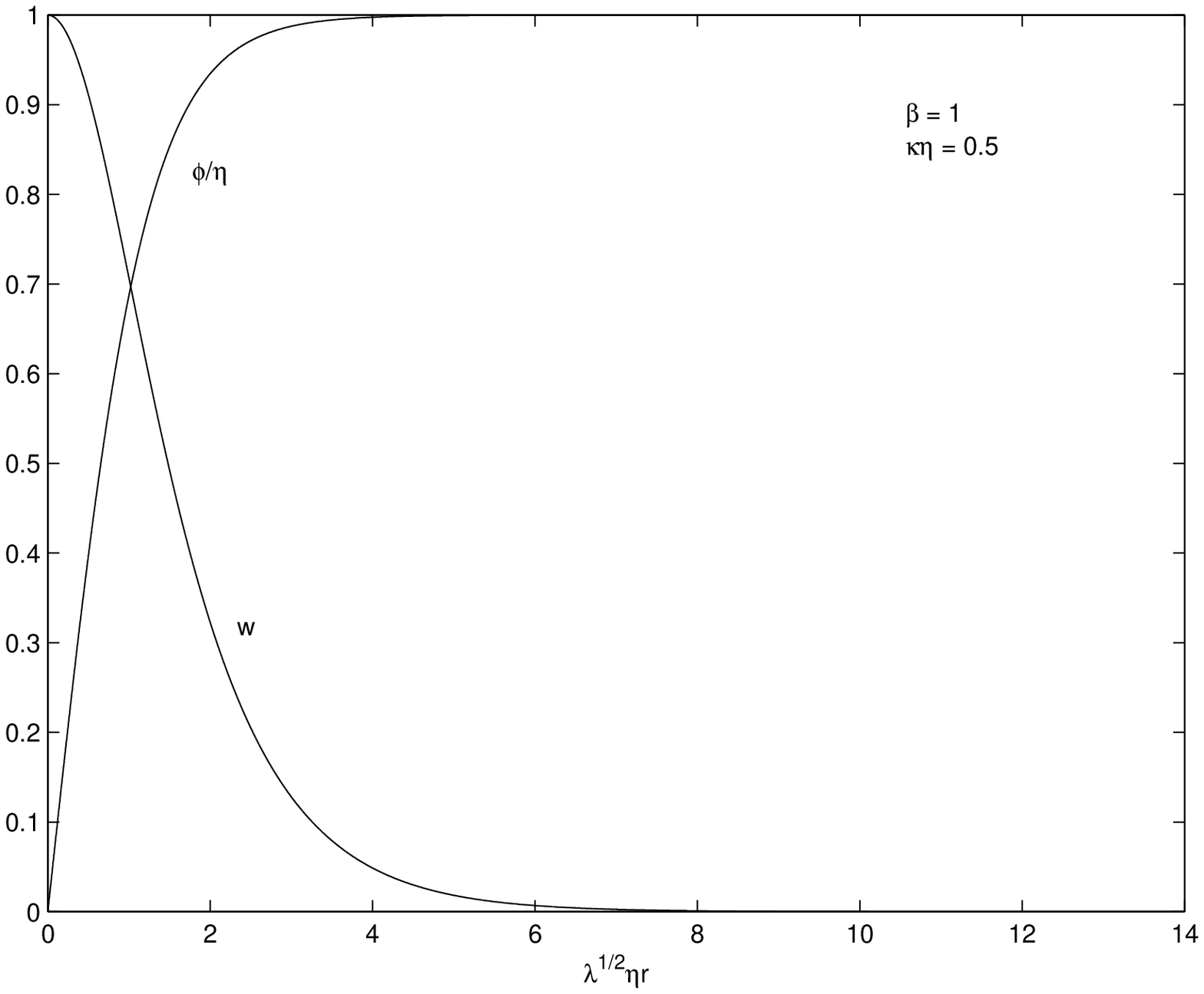,width=5in}
\end{center}
\vspace{0.5in}
\caption{
Flat-brane solutions: 
Scalar and gauge fields of a sub-critical monopole for
$\kappa\eta =0.5$ and $\beta\equiv \lambda/e^2 =1$.
The scalar field rapidly approaches $\eta$
and the gauge field approaches zero at large $r$.
}
\label{fig=H0regsfd}
\end{figure}

\begin{figure}
\begin{center}
\epsfig{file=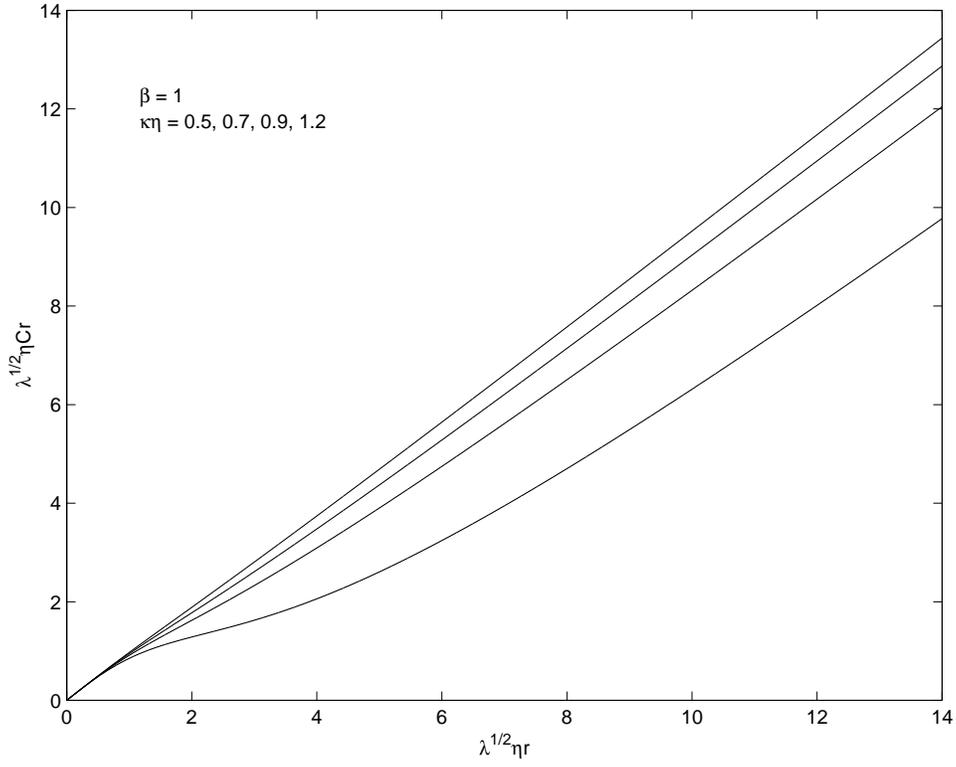,width=5in}
\end{center}
\vspace{0.5in}
\caption{
Flat-brane solutions:
The metric coefficient $Cr$ of sub-critical monopoles for $\beta=1$ and
$\kappa\eta = 0.5, 0.7, 0.9, 1.2$, from the top down.
As $\eta$ approaches the critical value, 
a `cigar'-like region develops near the core.
}
\label{fig=H0regCr}
\end{figure}

\begin{figure}
\begin{center}
\epsfig{file=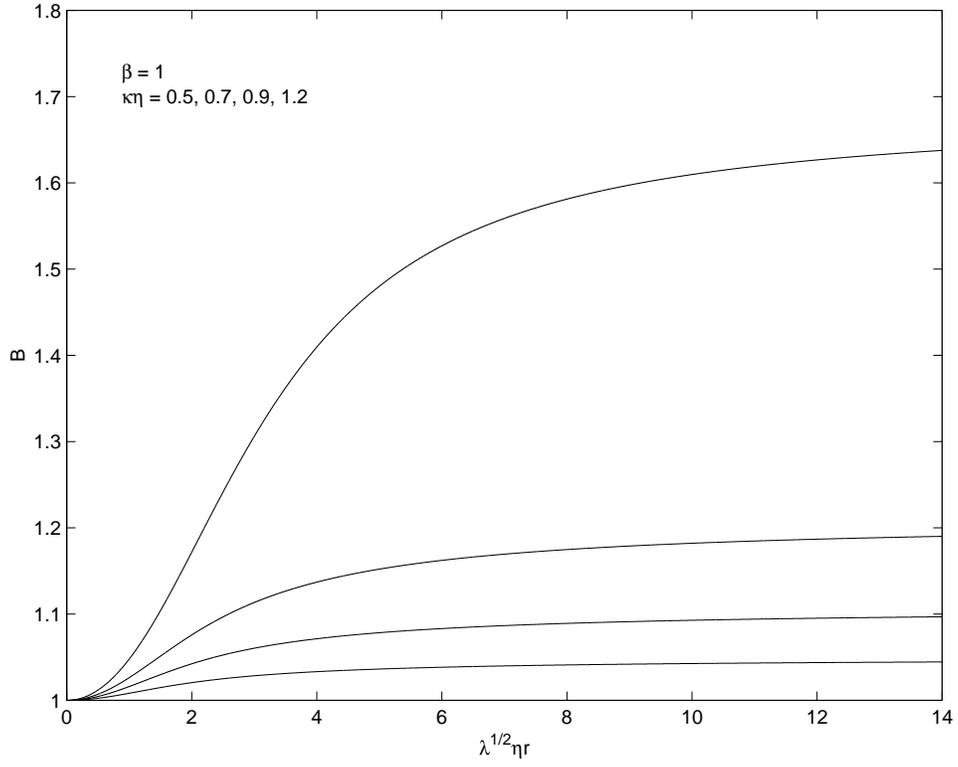,width=5in}
\end{center}
\vspace{0.5in}
\caption{
Flat-brane solutions:
The metric coefficient $B$ of sub-critical monopoles for $\beta=1$ and
$\kappa\eta = 0.5, 0.7, 0.9, 1.2$, from the bottom up.
}
\label{fig=H0regB}
\end{figure}

For a sufficiently small symmetry-breaking scale $\eta$, we find
regular, asymptotically flat solutions with $B(\infty)={\rm
const}$ and $C(\infty)=1$.  The fields $\phi$ and $w$ rapidly
approach their asymptotic values outside the monopole core; see
Fig.~\ref{fig=H0regsfd}.  As $\eta$ is increased, the character of the
solution gradually changes: the asymptotic value of $B(r)$ grows and a
`cylindrical' region develops near the core, where $C(r)r\approx {\rm
const}$.  This is illustrated in Figs.~\ref{fig=H0regCr} and
\ref{fig=H0regB}.

\begin{figure}
\begin{center}
\epsfig{file=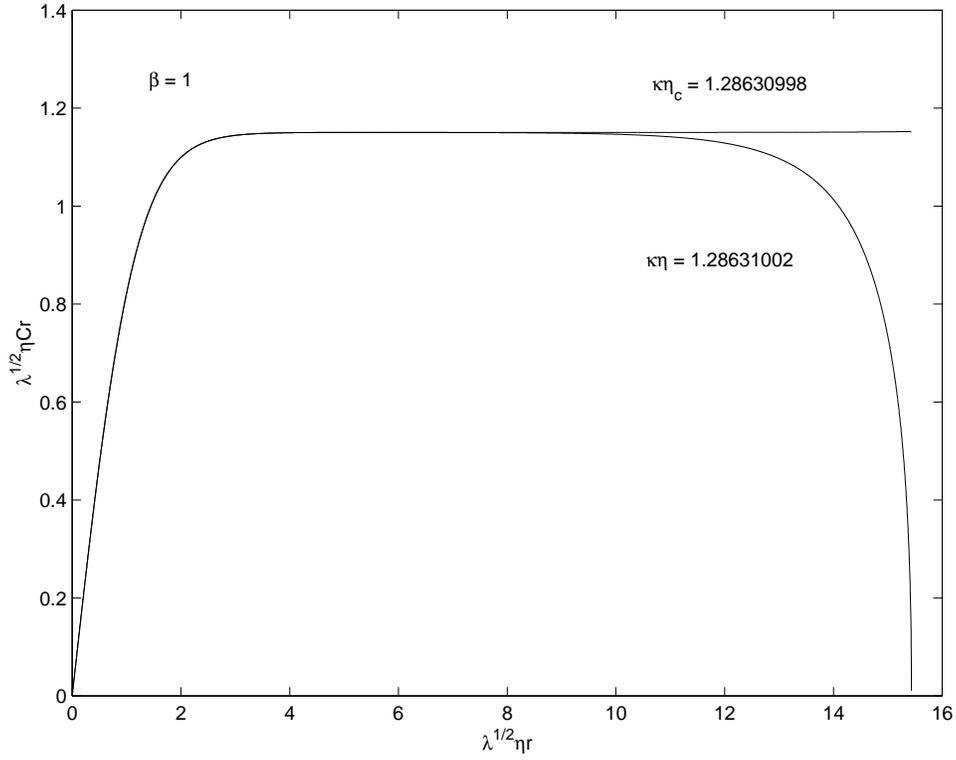,width=5in}
\end{center}
\vspace{0.5in}
\caption{
Flat-brane solutions:
The metric coefficient $Cr$ of a critical monopole,
$\kappa\eta_c = 1.28630998$, and of a super-critical
monopole, $\kappa\eta =  1.28631002$, for $\beta =1$.
For the critical monopole, $Cr$ is a constant, while
for the super-critical monopole it drops to zero at
the singularity.
}
\label{fig=H0singCr}
\end{figure}

\begin{figure}
\begin{center}
\epsfig{file=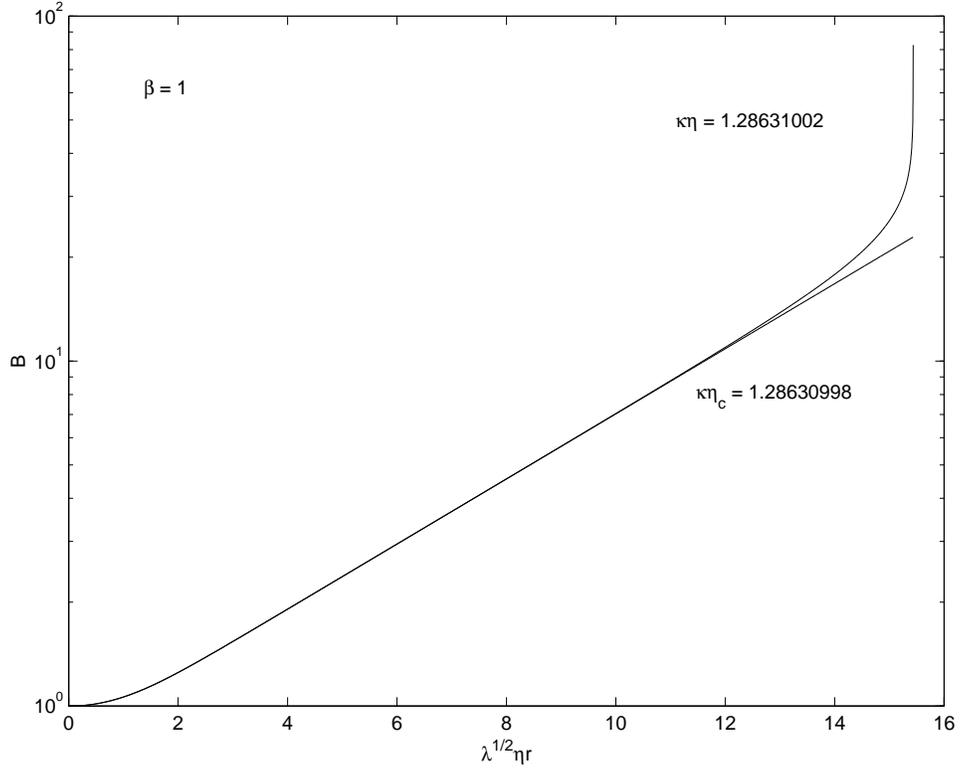,width=5in}
\end{center}
\vspace{0.5in}
\caption{
Flat-brane solutions:
The metric coefficient $B$ of a critical and a super-critical
monopole, for the same parameter values as in Fig.~\ref{fig=H0singCr}.
For the critical monopole, $B$ is an exponential function of $r$,
while for the super-critical monopole it diverges at the singularity.
}
\label{fig=H0singB}
\end{figure}

At some critical value $\eta=\eta_c$, the solution degenerates into a
`cigar', with $Cr$ approaching a constant at large $r$, and for $\eta
>\eta_c$ all static 
solutions are singular.  The singularity is at a finite value of $r$,
where $C(r)$ vanishes and $B(r)$ diverges (see
Figs.~\ref{fig=H0singCr} and \ref{fig=H0singB}).  This
singular behavior is opposite to that of a global monopole, for which
$C$ diverges and $B$ vanishes at the singular point \cite{GM}.  

\begin{figure}
\begin{center}
\epsfig{file=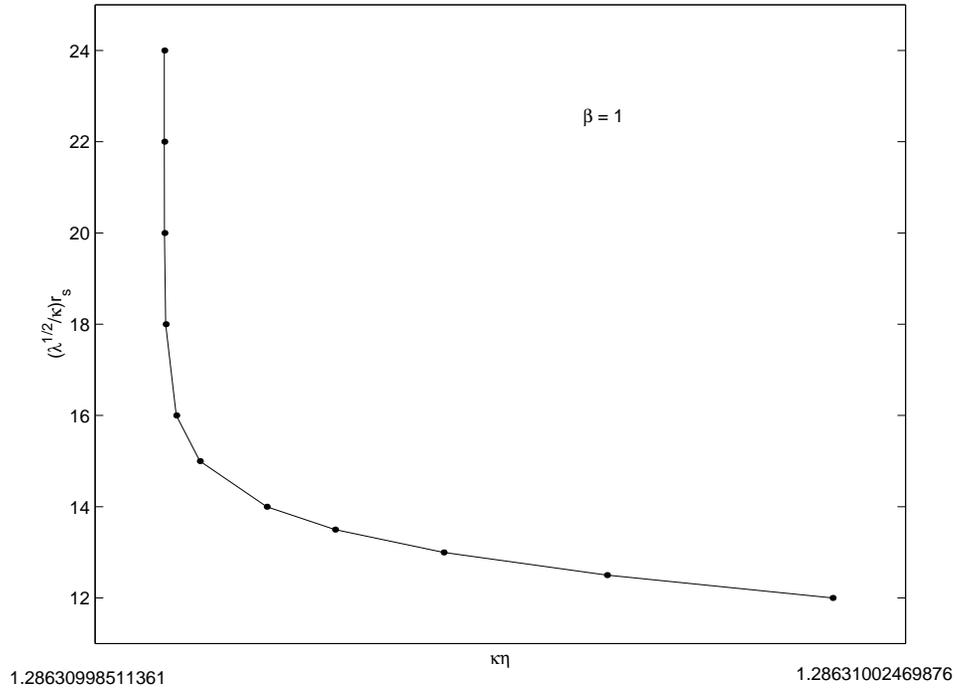,width=5in}
\end{center}
\vspace{0.5in}
\caption{
Location of the singularity $r_s$ vs. $\eta$ for $\beta=1$.
$r_s$ moves away from the core as $\eta$ is decreased and becomes
infinite at the critical value $\eta_c$.
}
\label{fig=rs}
\end{figure}

Figure~\ref{fig=rs} shows the location of the singularity, $r=r_s$, as a
function of $\eta$. As $\eta$ is decreased, the singularity moves away
from the core, and $r_s\to\infty$ in the limit $\eta\to\eta_c$.

The asymptotic form of the critical `cigar' solution can be found
analytically.  With the {\it ans\"atze} $Cr={\rm const}\equiv c_0$,
$\phi=\eta$, $w=0$, Einstein equations take the form
\beqr
-G^\mu_\mu &=& -3{B''\over B}-3\left({B'\over B}\right)^2
+{1\over c_0^2} = {\kappa^2\over 2e^2c_0^4}\,,\\
-G^r_r &=& -6\left({B'\over B}\right)^2+{1\over c_0^2}
= {\kappa^2\over 2e^2c_0^4}\,,\\
-G^{\theta_i}_{\theta_i} &=& -4{B''\over B}-6\left({B'\over B}\right)^2
= -{\kappa^2\over 2e^2c_0^4}\,,
\eeqr
and the solution is easily found,
\beq
B = b_0e^{{\sqrt{5}e \over 8\kappa}r}\,,\qquad
c_0 = {2 \over \sqrt{5}}{\kappa\over e}\,,
\label{eq=BCflat}
\eeq
where $b_0$ is an integration constant. The value of $\eta_c$ depends
on the parameter $\beta =\lambda/e^2$ and has to be determined
numerically.  The cigar solution in
Figs.~\ref{fig=H0singCr} and~\ref{fig=H0singB} is obtained for
$\beta=1$, in which case we find $\kappa\eta_c = 1.28630998$.

\begin{figure}
\begin{center}
\epsfig{file=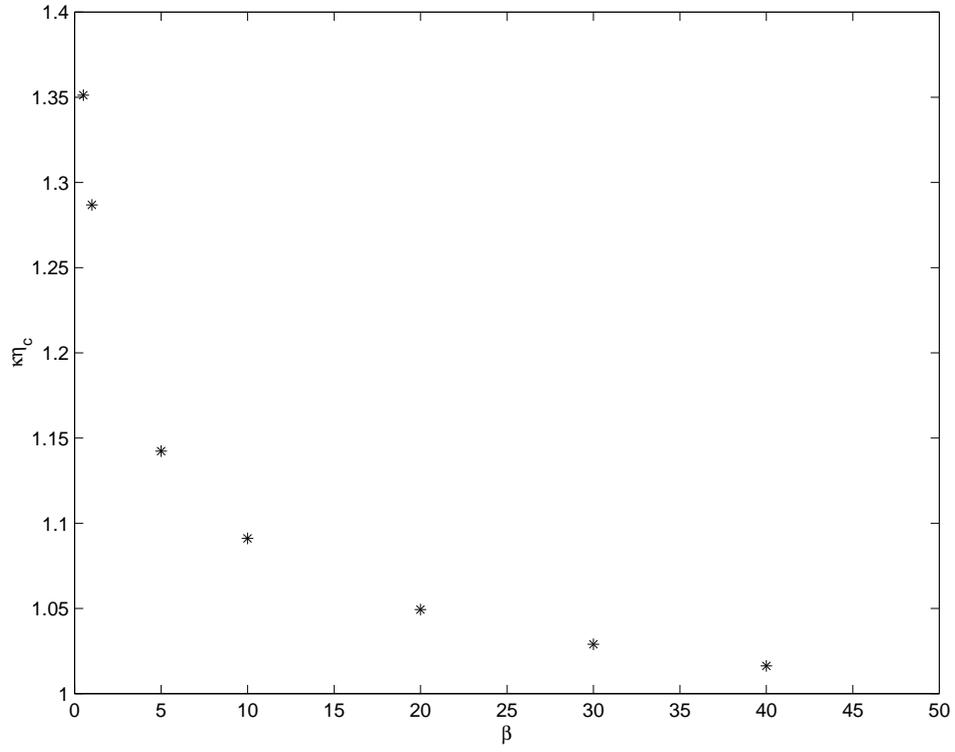,width=5in}
\end{center}
\vspace{0.5in}
\caption{
Flat-brane solutions:
The critical value $\kappa\eta_c$ for several values of 
$\beta =0.5,1,5,10,20,30,40$.
}
\label{fig=alphaeta}
\end{figure}

In Fig.~\ref{fig=alphaeta}, we show the critical value $\kappa\eta_c$
for several values of $\beta$.
As $\beta$ increases, $\kappa\eta_c$ decreases and approaches
$\kappa\eta_c=1$ in the global monopole limit $\beta\to\infty$.

\section{Inflating solutions}

\subsection{Sub-critical case: $\eta<\eta_c$}

We shall now discuss monopoles with inflating 4D worldsheets.  In this
case, the worldsheet metric is a 4D de Sitter space
(\ref{deSitter}), and the 4D curvature scalar is
$\bar{R}^{(4)}=12H^2$.

\begin{figure}
\begin{center}
\epsfig{file=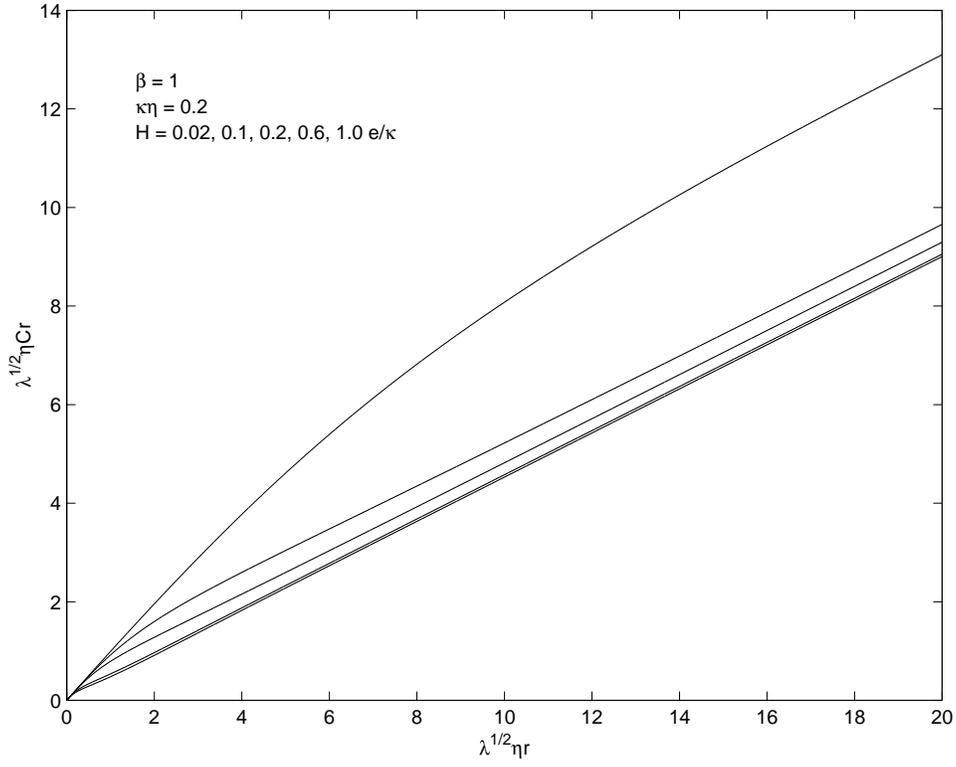,width=5in}
\end{center}
\vspace{0.5in}
\caption{
Inflating-brane solutions:
The metric coefficient $Cr$ of sub-critical monopoles for
$\kappa\eta = 0.2$ and $\beta =1$, with the expansion rate
$H=0.02, 0.1, 0.2, 0.6, 1.0 e/\kappa$ from the top down.
}
\label{fig=HregCr}
\end{figure}

\begin{figure}
\begin{center}
\epsfig{file=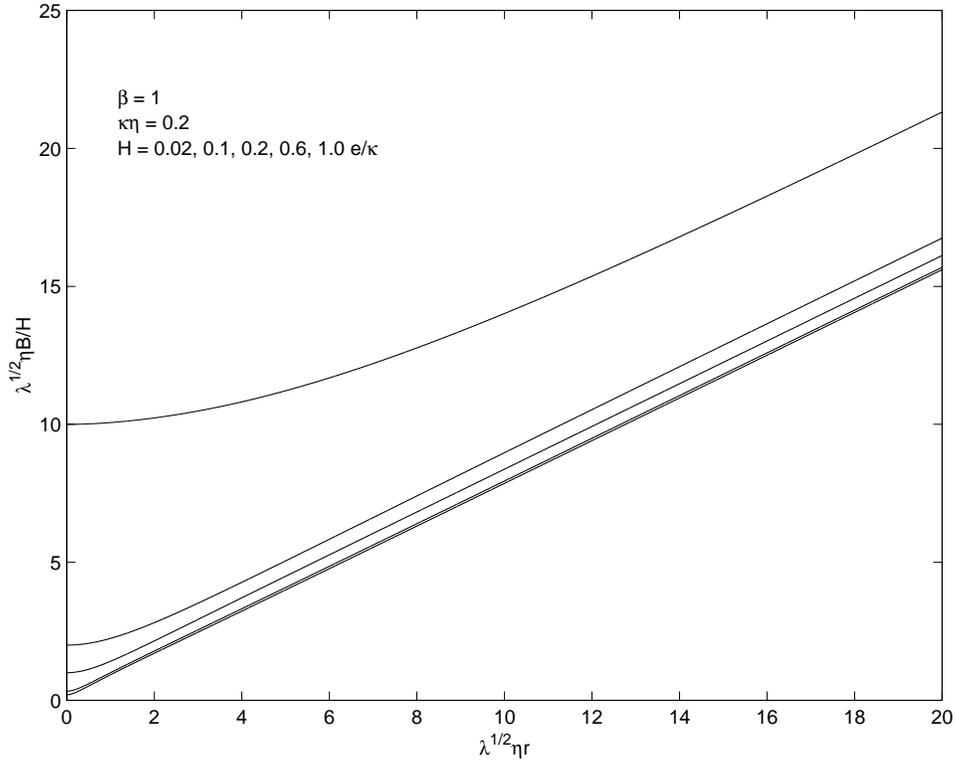,width=5in}
\end{center}
\vspace{0.5in}
\caption{
Inflating-brane solutions:
The metric coefficient $B$ of sub-critical monopoles for
$\kappa\eta = 0.2$ and $\beta =1$, with
$H=0.02, 0.1, 0.2, 0.6, 1.0 e/\kappa$ from the top down.
}
\label{fig=HregB}
\end{figure}

Let us first consider sub-critical inflating monopoles with $\eta$
significantly below $\eta_c$.  In this case, we found regular
solutions for a range of expansion rates $H$ spanning almost two
orders of magnitude.  All these solutions approach Gregory's
asymptotic form (\ref{Gregory}) at large $r$. We note also that, as
$H$ is increased, our solutions appear to approach Gregory's metric
(\ref{Gregory}) at all $r$ in the limit $H\to\infty$.  This is
illustrated in Figs.~\ref{fig=HregCr} and \ref{fig=HregB} for
$\kappa\eta=0.2$.

A peculiar feature of these solutions is that the expansion rate $H$
is not determined by the energy density of the brane, as in the case
of the ``usual'' 4D inflation, but is rather a free parameter.  Also
peculiar is the asymptotic geometry at large $r$, which exhibits a
deficit solid angle $\Delta\Omega=16\pi/5$, also independent of the
stress-energy of the brane. It thus appears that inflation is imposed
on the brane by the choice of boundary conditions.

\begin{figure}
\begin{center}
\epsfig{file=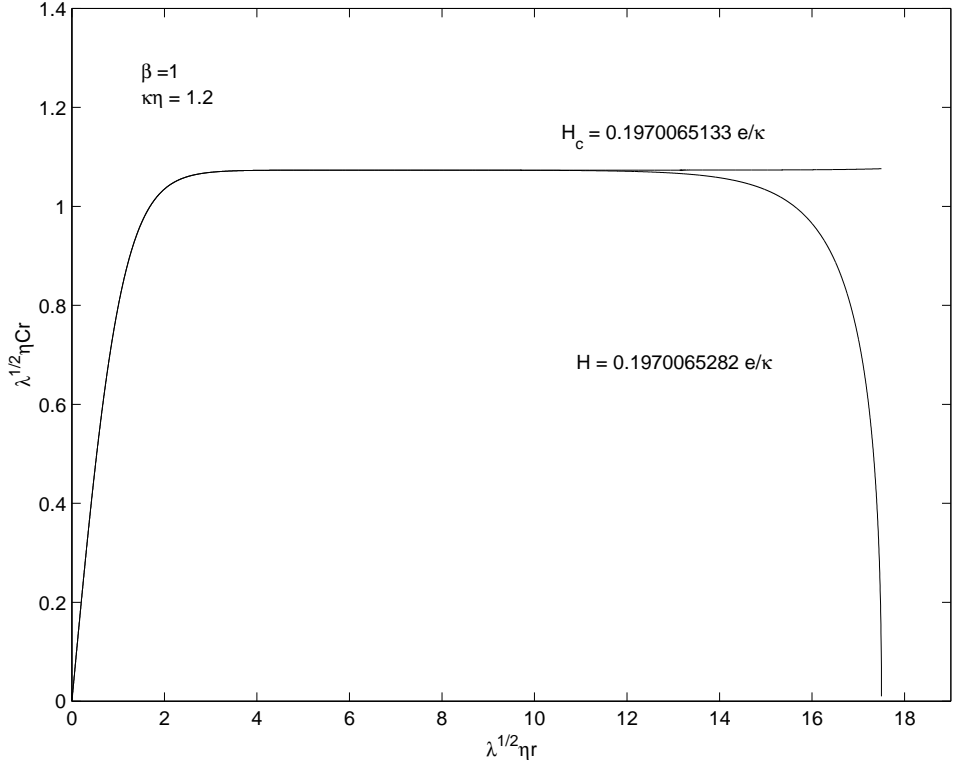,width=5in}
\end{center}
\vspace{0.5in}
\caption{
Inflating-brane solutions:
The metric coefficient $Cr$ of a sub-critical monopole
with $\kappa\eta =  1.2 <\kappa\eta_c$ and $\beta =1$, for two values
of the expansion rate $H$: the critical rate
$H_c=0.1970065133  e/\kappa$, and a super-critical
rate $H =0.1970065282 e/\kappa$.
}
\label{fig=HsingCr}
\end{figure}

\begin{figure}
\begin{center}
\epsfig{file=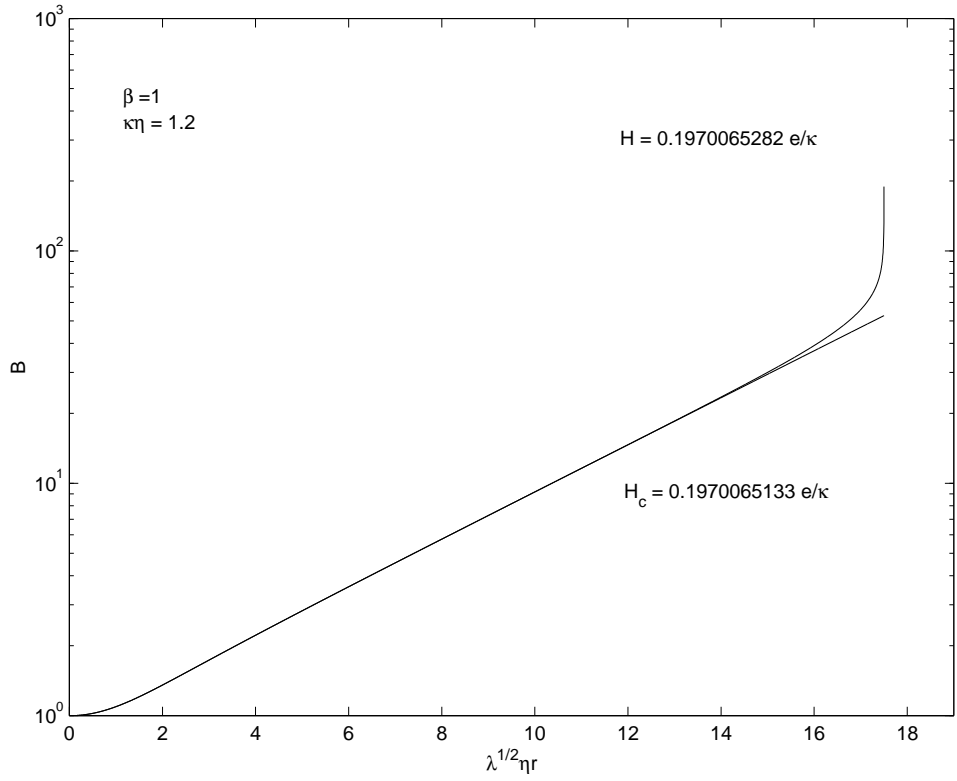,width=5in}
\end{center}
\vspace{0.5in}
\caption{
Inflating-brane solutions:
The metric coefficient $B$ of a sub-critical inflating monopole,
for the same parameter values as in Fig.~\ref{fig=HsingCr}.
}
\label{fig=HsingB}
\end{figure}

For larger values of $\eta$, $1/\kappa\lesssim \eta <\eta_c$, we find
a different pattern. As $H$ is increased, a cigar solution with
$Cr={\rm const}$ is obtained at some critical $H_c(\eta)$
\cite{foot1}. At still larger values of $H$, the solutions have a
curvature singularity of the type similar to the $H=0$ case (see
Figs.~\ref{fig=HsingCr} and
\ref{fig=HsingB}).

Based on this analysis, it appears reasonable to conjecture that the
physical solutions for sub-critical branes ($\eta<\eta_c$) are the
static regular solutions with $H=0$. The physical interpretation of
inflating sub-critical solutions is not clear to us, and we are
inclined to dismiss them as unphysical. We have not, therefore,
attempted to determine the dependence $H_c(\eta)$ in any detail.

\subsection{Super-critical case: $\eta>\eta_c$}

In the super-critical case, $\eta>\eta_c$, regular static solutions no
longer exist.  It is well known that the singularity of some static
defect solutions can be removed by relaxing the requirement of
staticity and allowing the defect worldsheet to inflate.  Examples are
domain walls~\cite{AV83,Ipser} and global
strings~\cite{Gregory,Santos} in $(3+1)$ dimensions.  In both of these
cases, the inflation rate $H$ is uniquely determined by the
requirement of regularity, and the inflating solutions can be
interpreted as true physical spacetimes for these sources.  The same
idea was applied to a 7D global monopole with a 3D core~\cite{GM}
(this corresponds to $e=0$, or $\beta=\infty$, in our model).  In
that paper, we found {\it (i)} that regular static solutions exist
only for $\eta<\eta_c=\kappa^{-1}$, and {\it (ii)} that nonsingular
super-critical solutions with $\eta>\eta_c$ exist only for inflating
branes.  In these solutions, the scalar field acquires an asymptotic
value which is somewhat lower than the symmetry-breaking scale,
$\phi=\phi_0<\eta$, the extra dimensions have a cigar geometry, and
we have argued that the value of $H$ is uniquely fixed by the
regularity requirement.

We have verified that regular solutions of the same type also exist for
super-critical gauge monopoles. The asymptotic form of these solutions
at large $r$ can be found analytically. With the {\it ans\"atze}
$\phi=\phi_0$, $w=w_0$, $\sqrt{\lambda}\eta Cr = C_0$, where $\phi_0$,
$w_0$ and $C_0$ are constants, the combined field equations yield the
following equation for the scalar field $\phi_0$: 
\beq
{4\kappa^2 \over 5\eta^2}\left({1\over \beta}-{1\over 2}\right)\phi_0^4
+{1\over \eta^2}\left({3\kappa^2\eta^2 \over 10}-{1\over \beta}
+{1\over 2}\right)\phi_0^2
+{\kappa^2\eta^2\over 10}-{1\over 2} = 0\,.
\eeq
The solution to this equation, as a function of $\eta$ and $\beta$,
is given by
\beq
\left( \phi_0^{\pm}\right)^2 = \left[-\left({3\kappa^2\eta^2 \over 10} -{1\over
\beta}+{1\over 2}\right)
\pm \sqrt{D}\right]
\left[ {8\kappa^2 \over 5}\left( {1\over \beta}-{1\over 2}\right) \right]^{-1},
\label{eq=phi0}
\eeq
where
\beq
D=\left({3\kappa^2\eta^2 \over 10} -{1\over \beta}+{1\over 2}\right)^2
-{16\kappa^2\eta^2\over 5} \left({1\over\beta}-{1\over 2}\right)
\left({\kappa^2\eta^2 \over 10} -{1\over 2} \right)\,.
\label{eq=D}
\eeq
With this expression for the scalar field, the other fields are
\beqr
\sqrt{\lambda}\eta B &=& {H\over \sqrt{K}}sinh(\sqrt{\lambda}\eta\sqrt{K}r)\quad (\text{for } K>0)\,,
\label{eq=Bshcigar}\\
\sqrt{\lambda}\eta B &=& {H\over \sqrt{-K}}\sin(\sqrt{\lambda}\eta\sqrt{-K}r)\quad (\text{for } K<0)\,,
\label{eq=Bscigar}\\
C_0^2 &=& \left[ {1\over 2}+\left({1\over\beta}-{1\over 2}\right)
{\phi_0^2\over \eta^2} \right]^{-1}\,,
\label{eq=Ccigar}\\
w_0^2 &=& {C_0^2 \over 2}\left( 1-{\phi_0^2\over \eta^2}\right) \,,
\label{eq=wcigar}
\eeqr
where
\beq
K = {\kappa^2\phi_0^2 -1 \over 4C_0^2}\,.
\label{eq=K}
\eeq
Note that it follows from Eq.~(\ref{eq=wcigar}) that $\phi_0<\eta$,
and it follows from (\ref{eq=K}) that $\phi_0>\kappa^{-1}$ for $K>0$
solutions and $\phi_0 <\kappa^{-1}$ for $K<0$ solutions.

In the asymptotic solutions~(\ref{eq=Bshcigar}) and~(\ref{eq=Bscigar}), 
the expansion rate $H$ can be
absorbed by rescaling the time coordinate $t$ in Eq.~(\ref{deSitter});
then the solution takes the form
\beqr
(\lambda\eta^2) ds^2 &=& K^{-1}( sinh^2\chi d\bar{s}_+^2 + d\chi^2)
+ C_0^2d\Omega_2^2\quad (\text{for } K>0)\,,
\label{itsaso21}\\
{} &=& |K|^{-1}( \sin^2\chi d\bar{s}_+^2 + d\chi^2)
+ C_0^2d\Omega_2^2\quad (\text{for } K<0)\,,
\label{itsaso2}
\eeqr
where $\chi=\sqrt{\lambda}\eta \sqrt{|K|}r$ and $d\bar{s}_+^2$ is the 4D de
Sitter metric with $H=1$. Note that, although $H$ drops out of the asymptotic
solution, it is a meaningful parameter for the full spacetime.  With
our boundary conditions at $r=0$, it gives the inflation rate in the
core of the monopole.

For $K<0$, the above metric is essentially the same as the one found
in Ref.~\cite{Olasagasti} as a solution for a global defect with $\phi
=\eta$ and a positive bulk cosmological constant.  It was also
obtained in Ref.~\cite{GM} as a cigar solution for a global monopole
with $\phi_0<\eta$. In the latter case, the role of the cosmological
constant is played by the scalar field potential, $V(\phi_0)>0$. For a
gauge monopole, there is in addition the gauge field energy density,
which approaches a constant at large $r$.

The solution~(\ref{itsaso2}) has an apparent singularity at
$\chi=\pi$, but as it was noted in Ref.~\cite{Olasagasti}, the first two
terms in Eq.~(\ref{itsaso2}) describe a 5D de Sitter space.  The 5D
inflation rate is $H_5=\sqrt{\lambda}\eta \sqrt{|K|}$, and the surface
$\chi=\pi$ corresponds to the de Sitter horizon.  This shows that the
space outside the monopole core inflates not only along the 4D
worldsheet, but also in one of the transverse directions, while two of
the extra dimensions remain compactified in a sphere.  The effective
5D cosmological constant is
\beq
\Lambda_5 = -6\lambda\eta^2 K,
\eeq
so the geometry of the uncompactified dimensions is flat for $K=0$ and
anti-de Sitter for $K>0$.

\subsection{The parameter space of asymptotic solutions}

We found in Ref.~\cite{GM} that in the global monopole case,
$\eta_c=1/\kappa$ and solutions of the form (\ref{itsaso2}) exist only in
the range $1\leq \kappa^2\eta^2 < 5$. For a gauge monopole, we also
find that asymptotic cigar solutions exist only in a restricted
parameter space of $\eta$ and $\beta$. The conditions for a solution
to exist are $D\geq 0$, $\phi_0^2>0$, $C_0^2 >0$, and $w_0^2>0$. 

\begin{figure}
\begin{center}
\epsfig{file=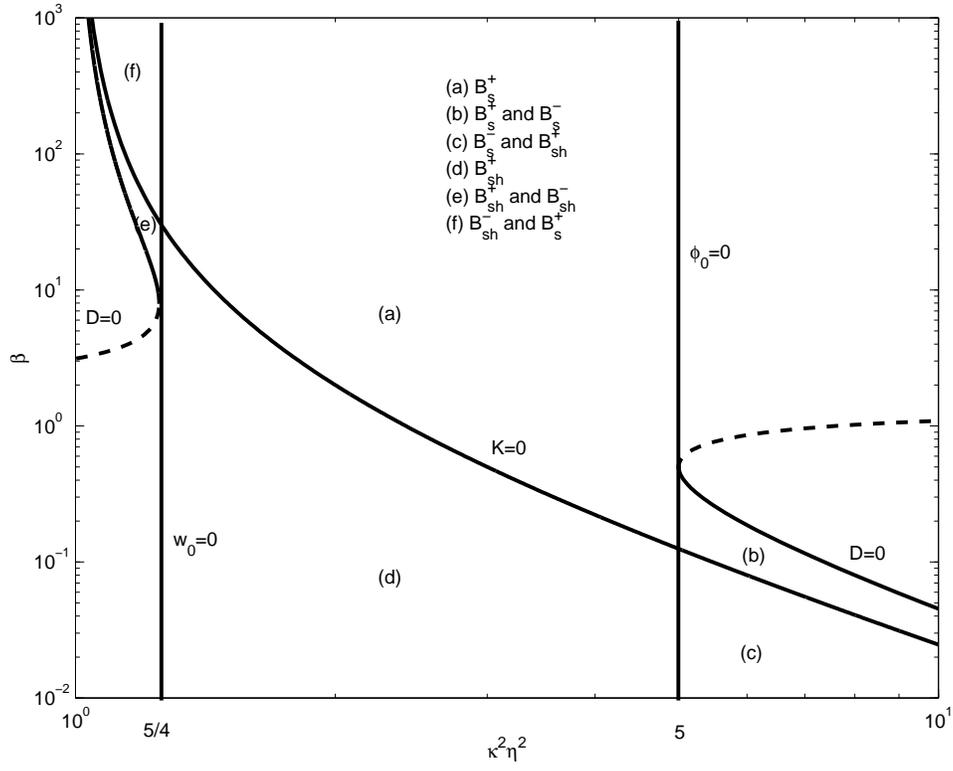,width=5in}
\end{center}
\vspace{0.5in}
\caption{
The parameter space of inflating cigar solutions.
}
\label{fig=paraspace}
\end{figure}

The $\eta-\beta$ parameter space is partitioned by the lines $D=0$,
$\phi_0=0$, $C_0=0$, $w_0=0$, and $K=0$, as shown in
Fig.~\ref{fig=paraspace}. The vertical straight lines, $\kappa^2\eta^2
=5$ and $\kappa^2\eta^2=5/4$ correspond to $\phi_0=0$ and $w_0=0$,
respectively. The curve $D=0$ is split into two sectors. The right
sector tangents the $\kappa^2\eta^2=5$ line at $\beta=1/2$, while the
left sector tangents the $\kappa^2\eta^2=5/4$ line at $\beta=8$ and
asymptotes to $\kappa^2\eta^2=1$ as $\beta\to\infty$. Finally, the
curve $K=0$ is given by
\beq
\beta = {2\over (\kappa^2\eta^2 - 1)^2}\,
\eeq
and also asymptotes $\kappa^2\eta^2=1$ at large $\beta$.
 
The parameter space is thus divided into several domains with
different types of solutions. In the figure, the domains are marked
(a)-(f) and the corresponding solutions are: (a) $B_s^+$, (b) $B_s^+$
and $B_s^-$, (c) $B_s^-$ and $B_{sh}^+$, (d) $B_{sh}^+$, (e)
$B_{sh}^+$ and $B_{sh}^-$, (f) $B_{sh}^-$ and $B_s^+$. The
superscripts $\pm$ indicate the choice of sign in the expression
(\ref{eq=phi0}) for $\phi_0$, and the subscripts indicate whether the
corresponding solution for $B$ is a sine function ($K<0$) or a sinh
function ($K>0$).

\subsection{Nonsingular super-critical solutions}

By analogy with the global monopole case, one can expect that, for a
given value of $\beta$ and $\eta>\eta_c(\beta)$, there is a single
value of $H$ that gives a nonsingular solution.  We would then
interpret that solution as the physical solution describing the
spacetime of a super-critical gauge monopole. Of course, it is
impossible to find the precise value of $H$ numerically, and all
super-critical global monopole solutions we considered in Ref.~\cite{GM}
eventually developed a singularity. By adjusting the value of $H$, we
were able to shift the onset of the singularity to larger and larger
radii, so that the solution got closer and closer to the analytic
asymptotic solution, presumably giving a better and better
approximation to the nonsingular solution.

We tried to use the same strategy for gauge monopoles. We found,
however, that numerical instabilities here are much more severe than
in the global case. As we tried to extend the solutions from the
origin towards larger values of $r$, they became unstable well before
we got close to the singularity. We could not, therefore, get any
information about the value of $H$ that yields the nonsingular
solution.

\begin{figure}
\begin{center}
\epsfig{file=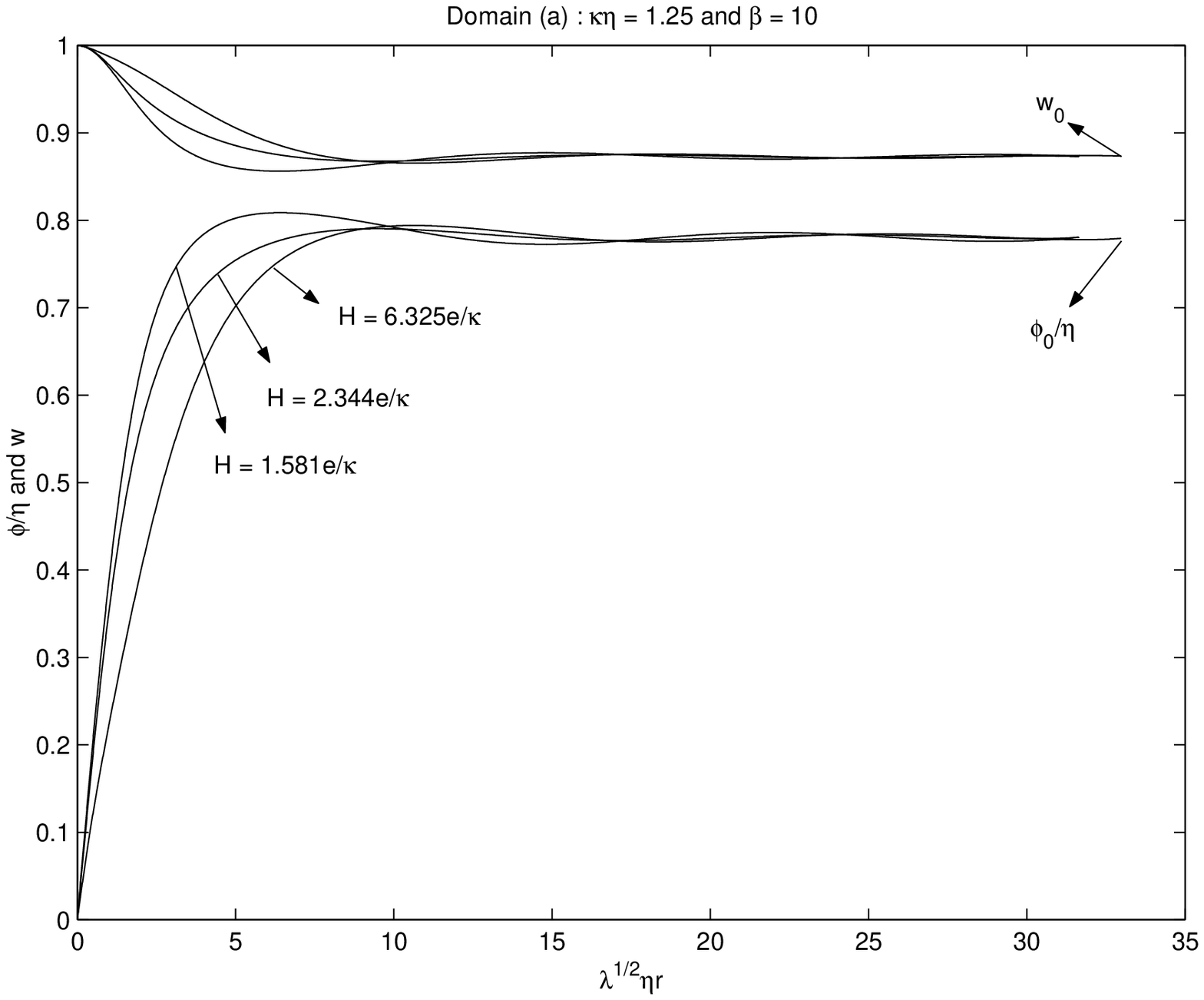,width=5in}
\end{center}
\vspace{0.5in}
\caption{
Inflating-brane solutions in domain (a): 
Scalar and gauge fields for $\kappa\eta =1.25$, $\beta =10$
and $H=1.581,2.344,6.325 e/\kappa$.
}
\label{fig=Asvfd}
\end{figure}

\begin{figure}
\begin{center}
\epsfig{file=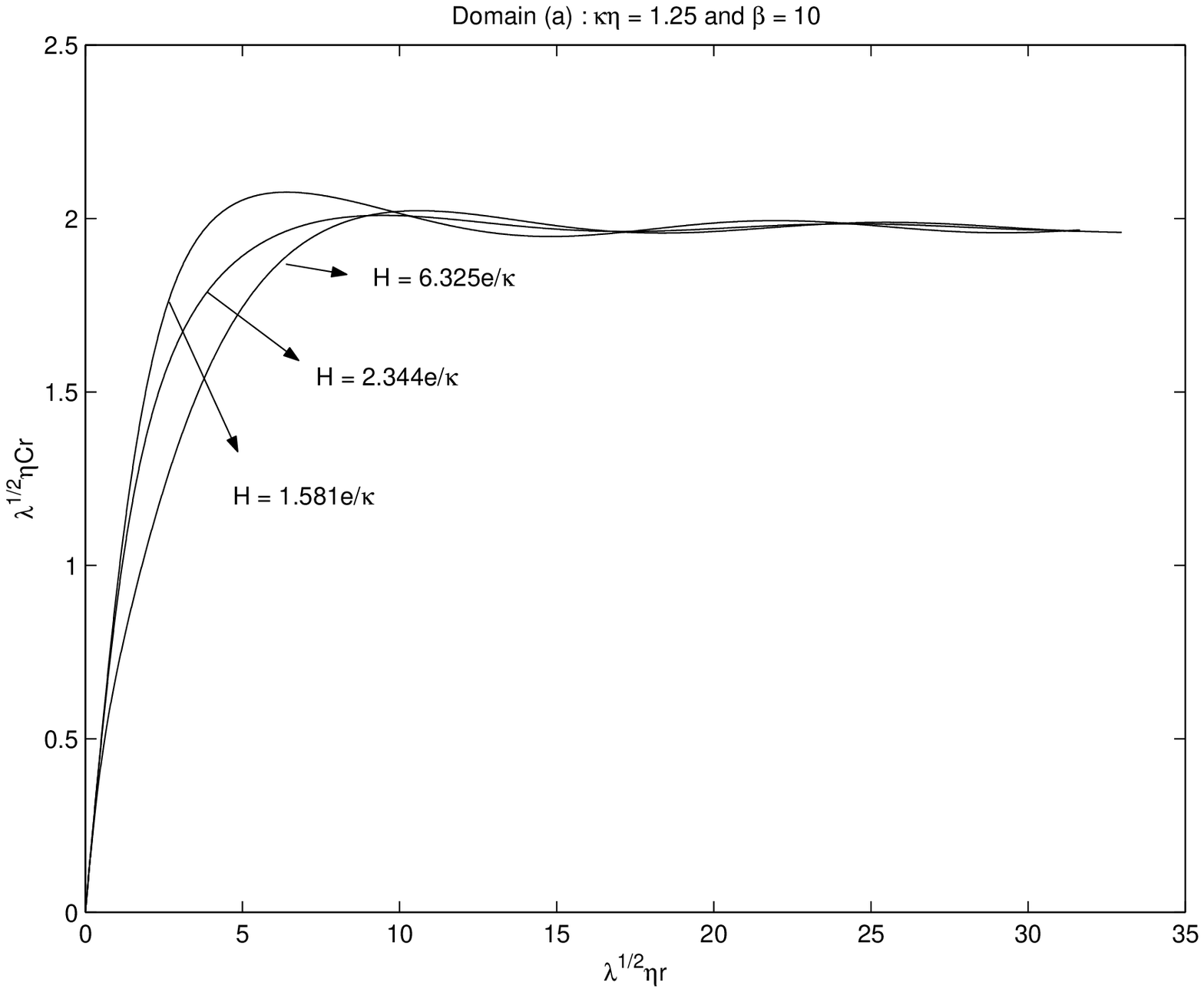,width=5in}
\end{center}
\vspace{0.5in}
\caption{
Inflating-brane solutions in domain (a): 
The metric coefficient $Cr$ for the same parameters as in
Fig.~\ref{fig=Asvfd}.
}
\label{fig=ACr}
\end{figure}

\begin{figure}
\begin{center}
\epsfig{file=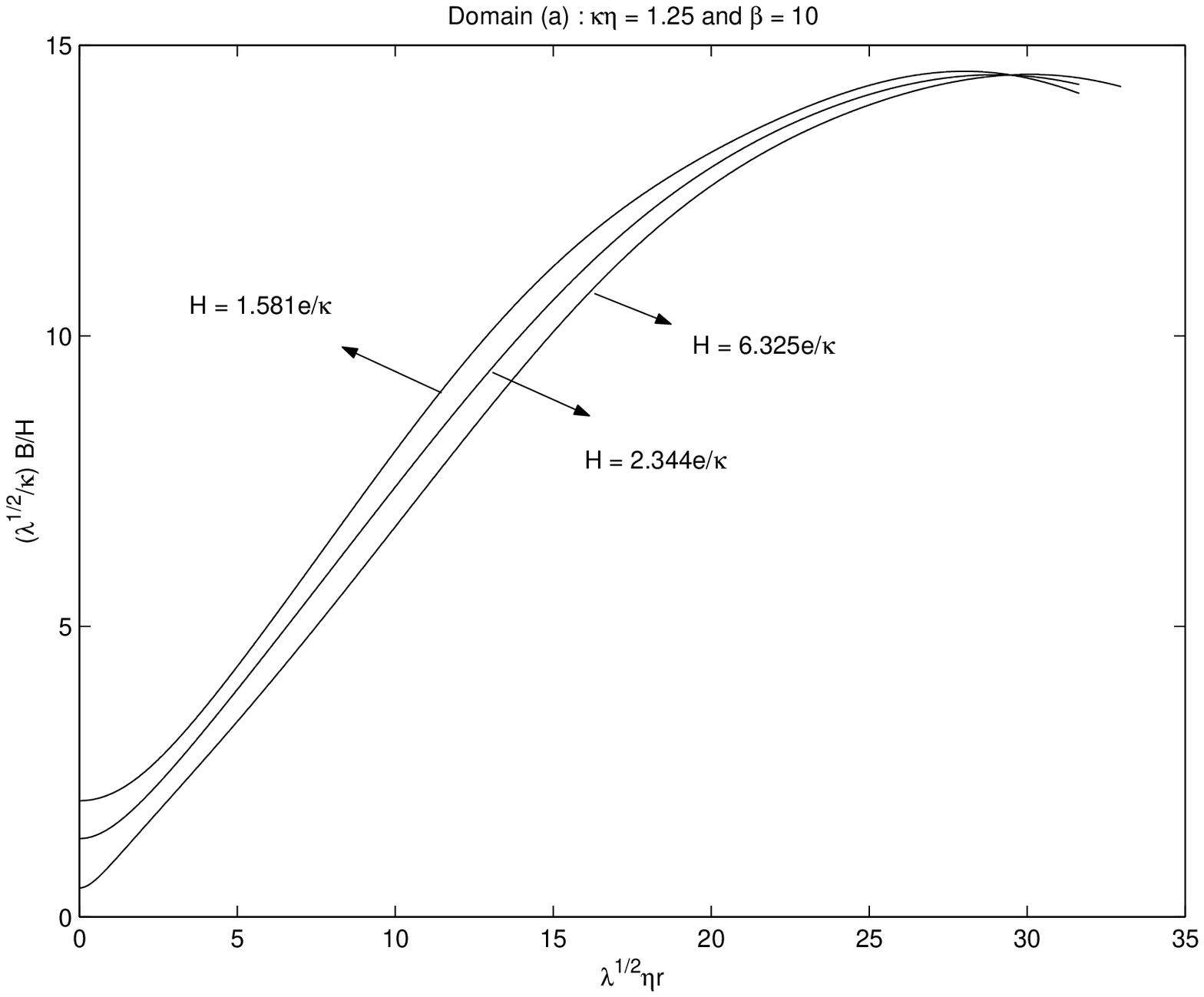,width=5in}
\end{center}
\vspace{0.5in}
\caption{
Inflating-brane solutions in domain (a): 
The metric coefficient $B$ for the same parameters as in
Fig.~\ref{fig=Asvfd}.
}
\label{fig=AB}
\end{figure}

\begin{figure}
\begin{center}
\epsfig{file=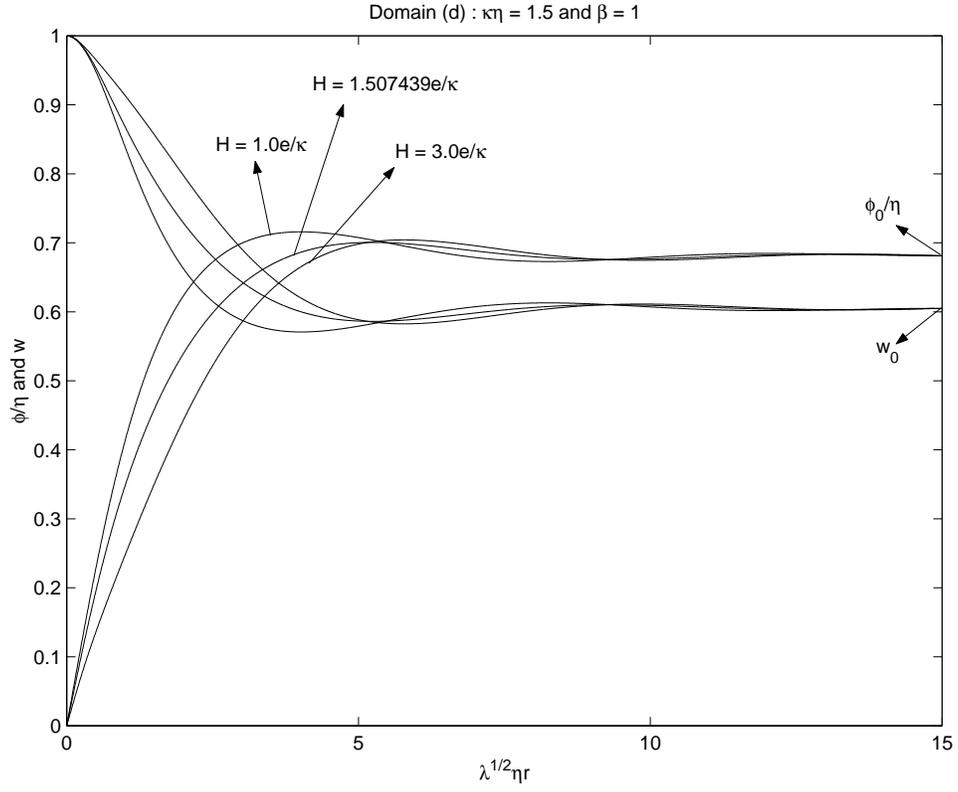,width=5in}
\end{center}
\vspace{0.5in}
\caption{
Inflating-brane solutions in domain (d): 
Scalar and gauge fields for $\kappa\eta =1.5$, $\beta =1$
and $H=1.0,1.507439,3.0 e/\kappa$.
}
\label{fig=Dsvfd}
\end{figure}

\begin{figure}
\begin{center}
\epsfig{file=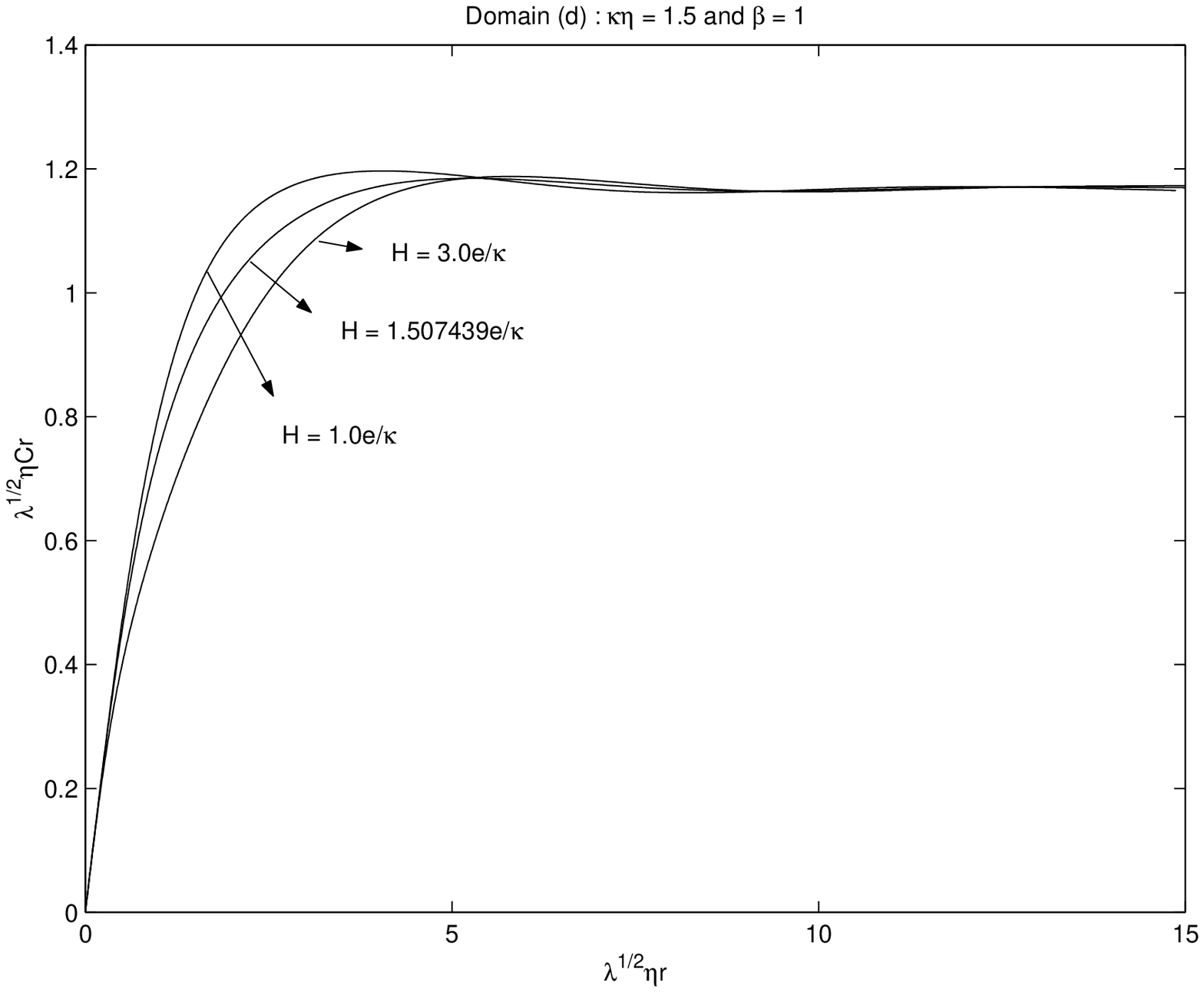,width=5in}
\end{center}
\vspace{0.5in}
\caption{
Inflating-brane solutions in domain (d): 
The metric coefficient $Cr$ for the same parameters as in
Fig.~\ref{fig=Dsvfd}.
}
\label{fig=DCr}
\end{figure}

\begin{figure}
\begin{center}
\epsfig{file=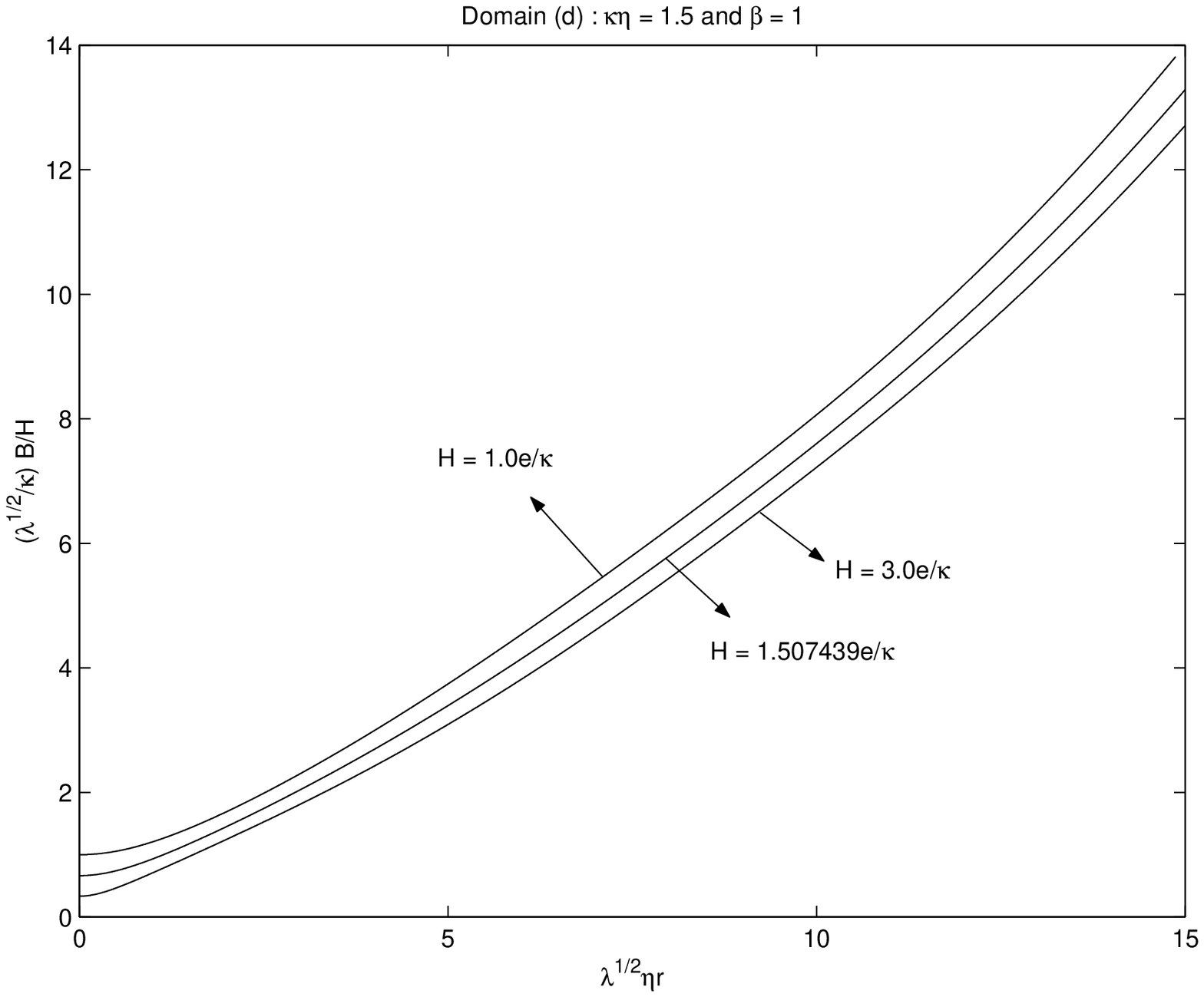,width=5in}
\end{center}
\vspace{0.5in}
\caption{
Inflating-brane solutions in domain (d): 
The metric coefficient $B$ for the same parameters as in
Fig.~\ref{fig=Dsvfd}.
}
\label{fig=DB}
\end{figure}

Sample numerical solutions with the parameter values in the domains
(a) and (d) are shown in Figs.~\ref{fig=Asvfd}-\ref{fig=DB} for
several values of $H$. The solutions are extended in $r$ as far as the
numerical accuracy allowed. All these solutions approach our
analytical asymptotic solutions at large $r$, indicating that
nonsingular super-critical inflating solutions do exist. However, our
numerical solutions exhibit this behavior for a wide range of $H$. We
cannot, therefore, conclude that the true nonsingular solution
corresponds to a single value of $H$, and if it does, our results do
not allow us to determine this value, even approximately. To make
further progress, one will have to find ways to significantly improve
the numerical accuracy. It may also be possible to prove the
uniqueness of the nonsingular solution using the dynamical systems
method employed in Refs.~\cite{Gregory,Santos}.

\section{Conclusions}

In this paper, we continued the investigation of the gravitational
field of higher-dimensional defects that we started in Ref.~\cite{GM}.
We have obtained numerical solutions of Einstein's, scalar, and gauge
field equations for a defect with a $(3+1)$-dimensional core, which
has a monopole-like field configuration in the extra three dimensions.
We have verified that, for symmetry-breaking scale $\eta$ below the
critical value $\eta_c$, the spacetime of the defect worldsheet is
flat, and the geometry of extra dimensions is asymptotically flat, in
agreement with the asymptotic solution obtained in Ref.~\cite{Gregory}.
For $\eta=\eta_c$, the extra dimensions degenerate into a `cigar', and
for $\eta>\eta_c$ all static solutions are singular. The critical
value $\eta_c$ depends on the ratio of scalar and gauge couplings,
$\beta=\lambda/e^2$, with $\eta_c$ approaching the Planck scale from
above as $\beta\to\infty$.

In the super-critical regime, $\eta>\eta_c$, we found numerical
solutions in which the defect core has the geometry of de Sitter space
inflating at some rate $H$, while extra dimensions have a cigar
geometry, with two of these dimensions compactified as a sphere of a
fixed radius.  At large distances from the core, the solutions have
a very simple form, and we found exact analytic solutions of the field
equations describing this asymptotic behavior.

The nature of super-critical monopole solutions is rather similar to
that of super-critical global monopoles that we discussed in
Ref.~\cite{GM}. There, our results indicated that, for a given $\eta$, there 
is a unique value of $H$ that gives a nonsingular solution, and that
$H$ is a growing function of $\eta$. This behavior is opposite to that
required in the Dvali, Gabadadze and Shifman (DGS) scenario for
solving the cosmological constant problem. One of our goals in the
present paper has been to find whether or not the situation is
different for gauge monopoles. Unfortunately, we were unable to check
this directly, since the numerical instabilities did not allow us to
determine the dependence of $H$ on $\eta$. In fact, our numerical
accuracy was not sufficient to conclude that the value of $H$ is fixed
(or strongly constrained) by fixing $\eta$ and requiring regularity of
the metric.

We do however have some indirect evidence indicating that
higher-dimensional magnetic monopoles do not help to solve the
cosmological constant problem. The DGS conjecture was originally based
on the assumption that in singular static monopole solutions~\cite{Charmousis}, 
the distance $r_s$ from the monopole center to the singularity is a
growing function of $\eta$. Then, in the inflating solutions, where
the singularity is replaced by a horizon, one expects that the
expansion rate $H\sim r_s^{-1}$ decreases with $\eta$. However, our
analysis of static solutions shows that $r_s$ gets smaller as $\eta$
is increased for magnetic monopoles, just as it does for global
monopoles. This suggests that the DGS scenario cannot be implemented
using inflating defect solutions.

\acknowledgements

We are grateful to Christos Charmousis, Ruth Gregory and Gia Dvali for useful
discussions.  The work of A.V. was supported in part by the National
Science Foundation.

\clearpage

\clearpage

\clearpage

\clearpage

\clearpage

\clearpage

\clearpage

\clearpage

\clearpage

\clearpage

\clearpage

\clearpage

\clearpage

\clearpage

\clearpage

\clearpage

\clearpage

\clearpage


\begin{references}


\bibitem{Rubakov}
V. Rubakov and M. Shaposhnikov,
Phys. Lett. {\bf 125B}, 139 (1983);
{\bf 125B}, 136 (1983).
\bibitem{Arkani}
N. Arkani-Hamed, S. Dimopoulos, and G. Dvali, Phys. Lett. B {\bf 429},
263 (1998); Phys. Rev. D {\bf 59}, 086004 (1999);
I. Antoniadis, N. Arkani-Hamed, S. Dimopoulos, and G. Dvali,
Phys. Lett. B {\bf 436}, 257 (1998).
\bibitem{RS}
L. Randall and R. Sundrum, Phys. Rev. Lett. {\bf 83}, 4690 (1999).
\bibitem{Wall}
G. Dvali and M. Shifman,
Phys. Lett. B {\bf 396}, 64 (1997).
\bibitem{String}
A. Cohen and D. Kaplan,
Phys. Lett. B {\bf 470}, 52 (1999);
R. Gregory,
Phys. Rev. Lett. {\bf 84}, 2564 (2000);
T. Gherghetta and M. Shaposhnikov,
Phys. Rev. Lett. {\bf 85}, 240 (2000).
\bibitem{Olasagasti}
I. Olasagasti and A. Vilenkin,
Phys. Rev. D {\bf 62}, 044014 (2000).
\bibitem{Monopole}
T. Gherghetta, E. Roessl, and M. Shaposhnikov,
Phys. Lett. B {\bf 491}, 353 (2000);
K. Benson and I. Cho,
Phys. Rev. D {\bf 64}, 065026 (2001);
E. Roessl and M. Shaposhnikov,
{\it ibid.} {\bf 66}, 084008 (2002);
K. Bronnikov and B. Meierovich, 
J. Exp. Theor. Phys. {\bf 97}, 1 (2003);
I. Olasagasti and K. Tamvakis,
hep-th/0303096.
\bibitem{Texture}
I. Cho,
Phys. Rev. D {\bf 67}, 065014 (2003).
\bibitem{Gregorypbrane}
R. Gregory,
Nucl. Phys. {\bf B467}, 159 (1996). 
\bibitem{GM}
I. Cho and A. Vilenkin,
Phys. Rev. D {\bf 68}, 025013 (2003). 
\bibitem{AV81}
A. Vilenkin, Phys. Rev. {\bf D23}, 852 (1981).
\bibitem{stringsing}
A. Cohen and D. Kaplan, Phys. Lett. B {\bf 215}, 67 (1988);
R. Gregory, Phys. Lett. B {\bf 215}, 663 (1988).
\bibitem{AV83}
A. Vilenkin, Phys. Lett. {\bf 133B}, 137 (1983).
\bibitem{Gregory}
R. Gregory,
Phys. Rev. D {\bf 54}, 4955 (1996).
\bibitem{V94}
A. Vilenkin,
Phys. Rev. Lett. {\bf 72}, 3137 (1994).
\bibitem{DGS}
G. Dvali, G. Gabadadze, and M. Shifman, Phys. Rev. D {\bf 67}, 044020 (2003).
\bibitem{Gregorypinf}
R. Gregory,
JHEP {\bf 0306}, 041 (2003). 
\bibitem{numerics}
In order to solve the equations numerically,
we used the ``relaxation method'' for the scalar and gauge
field equations, and
the ``shooting method'' for Einstein equations.
\bibitem{foot1}
The asymptotic form of the critical cigar solutions can be obtained 
analytically. With the
ansatz $\phi=\eta$, $w=0$, $Cr  ={\rm const}$, 
Einstein equations give:
\beqr
\sqrt{\lambda}\eta B &=& {H\over \sqrt{K}} 
sinh[\sqrt{\lambda}\eta\sqrt{K}(r-r_0)]
\label{eq=Bctcigar}\,,\\
\sqrt{\lambda}\eta Cr &\equiv & C_0 = {2\sqrt{\beta}\kappa\eta \over \sqrt{5}}
\label{eq=Cctcigar}\,,
\eeqr
where $r_0$ is an integration constant and
\beq
K= {5 \over 64 \beta\kappa^2\eta^2} \,.
\label{eq=Kctcigar}
\eeq
In this asymptotic solution, the expansion rate $H$ is a free
parameter, and the actual value of $H_c$ can only be determined
numerically. 
\bibitem{Ipser}
J. Ipser and P. Sikivie, Phys. Rev. D {\bf 30}, 712 (1984).
\bibitem{Santos}
R. Gregory and C. Santos, Class. Quant. Grav. {\bf 20}, 21 (2003).
\bibitem{Charmousis}
C. Charmousis, R. Emparan, and Ruth Gregory,
JHEP {\bf 0105}, 026 (2001).

\end{references}
\end{document}